\newcommand{\proton}{$^1$H}
\newcommand{\carbon}{$^{13}$C}

\newcommand{\degC}{$^\circ$C}
\newcommand{\Htwo}{H$_2$}

\newcommand{\Cat}{$\mathrm{[RuCp^*(CH_3CN)_3]PF_6}$}
\newcommand{\ADCA}{[1-\carbon]disodium acetylenedicarboxylate}

\newcommand\ket[1]{\left|#1\right>}
\newcommand\bra[1]{\left<#1\right|}
\documentclass[journal=ancham,manuscript=article,layout=manuscript]{achemso}
\setkeys{acs}{etalmode=truncate,maxauthors=20}
\usepackage{chemformula} 
\usepackage[T1]{fontenc} 
\usepackage[utf8]{inputenc}
\usepackage{soul}
\usepackage{mathptmx}
\usepackage{helvet}
\usepackage{xspace}
\usepackage{changebar}

\author{Sylwia J Barker}
\affiliation{School of Chemistry, University of Southampton, Southampton, United Kingdom}
\author{Laurynas Dagys}
\affiliation{School of Chemistry, University of Southampton, Southampton, United Kingdom}
\author{William Hale}
\affiliation{School of Chemistry, University of Southampton, Southampton, United Kingdom}
\altaffiliation{Department of Chemistry, University of Florida, Gainesville, USA}
\author{Barbara Ripka}
\affiliation{School of Chemistry, University of Southampton, Southampton, United Kingdom}
\author{James Eills}
\affiliation{Institute for Physics, Johannes Gutenberg University, D-55090 Mainz, Germany}
\altaffiliation{GSI Helmholtzzentrum für Schwerionenforschung GmbH, Helmholtz-Institut Mainz, 55128 Mainz, Germany}
\author{Manvendra Sharma}
\affiliation{School of Chemistry, University of Southampton, Southampton, United Kingdom}
\author{Malcolm H Levitt}
\affiliation{School of Chemistry, University of Southampton, Southampton, United Kingdom}
\author{Marcel Utz}
\affiliation{School of Chemistry, University of Southampton, Southampton, United Kingdom}
\email{marcel.utz@soton.ac.uk}

\title[Fumarate@Chip]
  {Direct Production of a Hyperpolarised Metabolite on a Microfluidic Chip}
\abbreviations{NMR,PHIP}

\begin{document}
\nochangebars

\begin{tocentry}
\centering
\includegraphics[width=7cm]{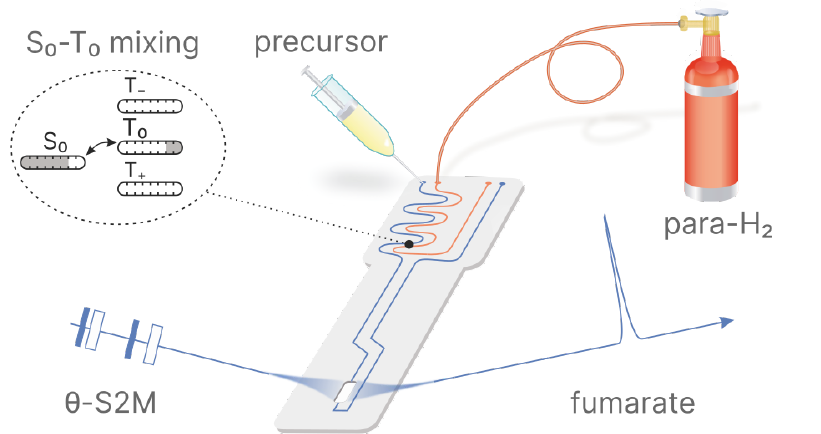}

\end{tocentry}

\begin{abstract}

Microfluidic systems hold great potential for the study of live microscopic cultures of cells, tissue samples, and small organisms. Integration of 
hyperpolarisation would enable quantitative studies of metabolism in such 
volume limited systems by high-resolution NMR spectroscopy. We demonstrate,
for the first time, the integrated generation and detection of a hyperpolarised
metabolite on a microfluidic chip. The metabolite [1-\textsuperscript{13}C]fumarate is produced
in a nuclear hyperpolarised form by (i) introducing para-enriched hydrogen into
the solution by diffusion through a polymer membrane, (ii) reaction with a substrate
in the presence of a ruthenium-based catalyst, and (iii) conversion of the 
singlet-polarised reaction product into a magnetised form by the application
of a radiofrequency pulse sequence, all on the same microfluidic chip. 
The microfluidic device delivers a continuous flow of hyperpolarised 
material at the 2.5~$\mu\text{L}/\text{min}$ scale, with a polarisation 
level of 4\%. We demonstrate two methods for mitigating singlet-triplet 
mixing effects which otherwise reduce the achieved polarisation level. 
\end{abstract}

\section{Introduction}
Nuclear magnetic resonance (NMR) is a versatile spectroscopic technique,
well-suited for noninvasively probing complex chemical systems and their
dynamic behaviour.  The sensitivity of NMR is limited by the polarisation of
nuclear spins, which is small in thermal equilibrium even at the largest
available magnetic fields. Hyperpolarisation methods such as
parahydrogen-induced polarisation (PHIP)~\cite{natterer_parahydrogen_1997,adams_reversible_2009,deninger_quantification_1999,bowers_transformation_1986,maly_dynamic_2008}
can produce much larger spin alignments in special cases, offering several
orders of magnitude enhancements in sensitivity. This is particularly
attractive in the context of microfluidic lab-on-a-chip (LoC) devices, where
sample volumes are typically of the order of nL to µL~\cite{Eills2021Sep}. Such LoC are versatile
platforms on which chemical and biological systems can be studied under
precisely controlled and reproducible conditions. LoC systems are commonly
used as scaffolds for cell~\cite{patra_time-resolved_2021,coluccio_microfluidic_2019,mehling_microfluidic_2014,du_microfluidics_2016,xiong_recent_2014}
and organ~\cite{wu_organ---chip_2020,jang_multi-layer_2010,stucki_lung---chip_2015,mandenius_conceptual_2018}
culture, providing valuable models for supporting the development of
diagnostics~\cite{kolluri_towards_2018,wu_lab--chip_2018}, therapies
\cite{wu_organ---chip_2020} and drug safety testing~\cite{jodat_human-derived_2018,cong_drug_2020},  but also for chemical reaction
monitoring~\cite{wu_rapid_2019}. While state-of-the-art micro-NMR probes can
provide \proton\ NMR detection sensitivities of around 1~nmol$\sqrt{\text s}$
for microliter-scale samples in a 14~T magnet at thermal equilibrium,~\cite{sharma_modular_2019}
this can be improved into the range of
$\text{pmol}\sqrt{\text s}$ by PHIP~\cite{eills_high-resolution_2019}. 
Like other hyperpolarisation methods, PHIP requires specific chemical processes and
spin manipulations to produce hyperpolarised species. LoC devices can be used
to implement some or all of these processes, thus offering the possibility
to integrate production and application of hyperpolarised species  in a single, compact
platform. 

PHIP is conventionally implemented by bubbling hydrogen gas enriched in 
the para spin isomer through a solution containing a suitable substrate and 
catalyst, either directly at high magnetic field (PASADENA experiments)~\cite{Bowers1987Sep} or
outside of the magnet at low ($\mathrm{\mu T}$) fields, followed by an adiabatic increase
of the magnetic field (ALTADENA experiments)~\cite{Pravica1988Apr}.
Such experiments are effective, but quite difficult
to repeat accurately. This complicates systematic studies of the interplay between
reaction kinetics and nuclear spin relaxation processes. As we have recently
shown, microfluidic implementation of PHIP at high field allows delivery of the
hydrogen gas by diffusion through a membrane, such that no bubbling is required~\cite{eills_high-resolution_2019}.
Experiments can therefore be carried out under continuous flow, with a stable stationary
level of hyperpolarisation established in the chip. This can be exploited for 
hyperpolarised multidimensional NMR experiments, which require superposition of many transients that
must maintain a high level of consistency.

In the following, we use the same approach to probe the formation of hyperpolarised
[1-\carbon]fumarate from \ADCA\ in an aqueous solution. To our
knowledge, this is the first demonstration of
PHIP-hyperpolarised metabolite  production 
in a microfluidic device.
Hyperpolarised fumarate is widely used as a 
contrast agent for in-vivo detection of necrosis~\cite{Gallagher2009Nov,Witney2010Oct,Bohndiek2010Dec,Clatworthy2012Aug,Mignion2014Feb,Miller2018Nov,Laustsen2020Jun,Knecht2021Mar,Wienands2021Jun,Eills2021Mar}.
While the current implementation is not yet ready for use with biological
systems due to the presence of the catalyst and other residues, the
stability of the microfluidic implementation allows systematic studies
of complex kinetic effects. 

In this work we generate and observe solutions of [1-\carbon]fumarate formed via trans-hydrogenative PHIP in a microfluidic
chip under continuous-flow conditions, performing the chemical reaction in one
part of the chip and NMR detection in another. The operation of this device has been discussed in detail elsewhere~\cite{eills_real-time_2019}.
Briefly, all of our experiments are performed inside of a high field NMR spectrometer where the reaction solution containing the precursor and catalyst is delivered to the chip via a syringe pump. Parahydrogen is delivered through a separate channel and diffuses through the PDMS membrane to dissolve into the precursor solution; hence the hydrogenation reaction takes place in the chip. 

\cbstart Microfluidic technology provides a convenient platform for studying hyperpolarised NMR experiments for the following reasons:~\cite{Eills2021Sep}
 \begin{enumerate}
 	\item The results are more reproducible since hydrogen is brought into
 		solution via diffusion through a membrane, which is less erratic than
 		bubbling or shaking~\cite{eills_high-resolution_2019,lehmkuhl2018continuous,Bordonali2019,roth2010continuous}.
 	\item The reaction kinetics and relaxation properties do not vary between
 		or during experiments since a steady-state can be established between
 		the rate of reaction and relaxation, and this can be finely tuned by,
 		e.g., varying the flow rates used~\cite{eills_high-resolution_2019,lehmkuhl2018continuous,Bordonali2019,roth2010continuous}.
 	\item The low volumes used in microfluidics (in this work a few
 		microlitres) makes it more practical to work with expensive or rare
 		samples. 
 	\item Since fresh reaction solution is continually provided to the
 		detection chamber, the samples do not need to be replaced between
 		experiments~\cite{eills_high-resolution_2019,Bordonali2019}.
 	\item Bringing the hyperpolarisation step close to the point of detection
 		minimises the signal losses due to relaxation.
 \end{enumerate}
 \cbend 
 
 Singlet-triplet mixing has been reported to hinder the achievable polarisation of [1-\carbon]fumarate at high field~\cite{rodin_constant-adiabaticity_2021,Wienands2021Jun,Dagys2020Aug}. Using our PHIP-on-a-chip system, we quantify how effectively two different RF pulse methods
mitigate the problem of ST mixing and support our finding with computational spin dynamics simulations. \cbstart We present quantitative data on the kinetics
and yield of [1-\carbon]fumarate from \ADCA\ in a microfluidic device. \cbend

\cbstart
\begin{figure*}[h!]
    \centering
    \includegraphics[width=\linewidth]{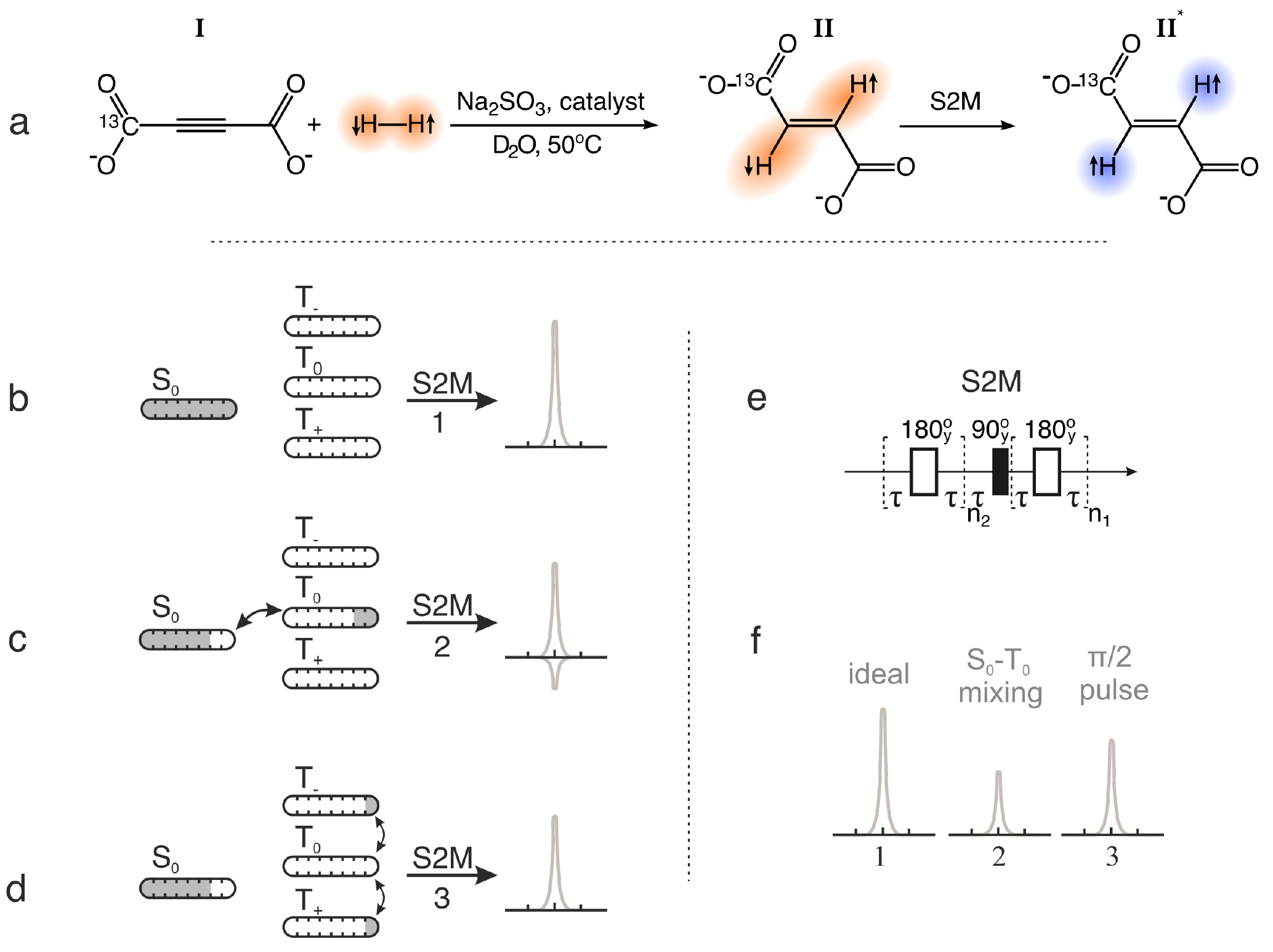}
    \caption{a) Scheme of the reaction investigated in this work. \ADCA\ labelled as molecule \textbf{I} reacts with parahydrogen in the presence of sodium sulfite and the catalyst \Cat\ in $\mathrm{D_2O}$ at 50\degC. The reaction results in a production of [1-$^{13}$C]disodium fumarate, molecule \textbf{II}, with the two protons in a singlet state. Application of the S2M pulse sequence converts the singlet state into a state that is magentic and hence observable, molecule  $\mathrm{\textbf{II}}^*$. b) Illustration of an ideal case where no ST mixing occurs; only $\ket{\mathrm{S_0}}$ state is populated. c) A case where ST mixing occurs leading to a leak of $\ket{\mathrm{S_0}}$ state population to the $\ket{\mathrm{T_0}}$ state. d) A case where the ST mixing is negated by applying a purge pulse prior the S2M, which distributes the population of $\ket{\mathrm{T_0}}$ state to $\ket{\mathrm{T_+}}$ and $\ket{\mathrm{T_-}}$ states. e) S2M pulse sequence converts the singlet order into observable hyperpolarised magnetisation. The optimal parameters for this molecular system are: $\tau=15.6$~ms, $n_2=14$, $n_1=7$. f) Predicted signal intensities for three different scenarios.}
    \label{fig:rxnscheme}
\end{figure*}
\cbend
  
\section{Background}
The reaction shown in \cbstart Fig.~\ref{fig:rxnscheme}a \cbend produces [1-\carbon]fumarate \textbf{II} by
hydrogenation of \ADCA\ \textbf{I} with parahydrogen in the presence of a ruthenium
catalyst.  The slight magnetic inequivalence due to the difference in \proton-\carbon\,
$J$-couplings
makes it possible to convert the singlet order into observable hyperpolarised
magnetisation through the use of RF pulse sequences. In this work we
use the singlet-to-magnetisation (S2M) pulse sequence for this task (see
Fig.~\ref{fig:rxnscheme}e), which is robust against field inhomogeneities in
contrast to alternative methods~\cite{Pileio17135}. This is important because magnetic field inhomogeneities are present in the chip due to differences in magnetic susceptibility of the chip and the solvent~\cite{Hale2018Sep}.  Applying this sequence after the
chemical reaction with parahydrogen results in high magnetisation of the two
protons giving rise to a hyperpolarised substance $\mathrm{\textbf{II}}^*$.

The polarisation that is generated on the target
molecules can be attenuated by singlet-triplet (ST) mixing
(sometimes called ST leakage)~\cite{kating_nuclear_1993}.  The hydrogen
molecules can form intermediate hydride species with the catalyst metal center,
where the two hydrogen atoms take up inequivalent positions, such that they
experience a chemical shift difference at high field.
If the lifetime of this
intermediate complex is long enough, there can be a significant leakage from the
\Htwo\ proton singlet state \cbstart ($\ket{S_0}$) \cbend to the central triplet state ($\ket{T_0}$), which
generally reduces the resulting PHIP signals.
\cite{bargon_nuclear_1993,kating_nuclear_1993,Dagys2021Oct}.
The S2M sequence converts both the $\ket{S_0}$ and the $\ket{T_0}$ states to
magnetisation, but with opposite phases. \cbstart The population of the $\ket{T_0}$ state therefore
\emph{reduces} the resulting NMR signal, as
illustrated in Fig.~\ref{fig:rxnscheme}b and c.\cbend\ This process sometimes gives rise to a
partially-negative line (PNL) in the \proton\ NMR
spectra~\cite{kiryutin_parahydrogen_2017}.
It is also known to occur in non-hydrogenative PHIP experiments, and has been noted to give rise
to `spontaneous' polarisation on the target
molecules~\cite{knecht_mechanism_2018}, although generally ST mixing is
undesirable.

Two methods have been shown to suppress ST mixing: spin locking on the \proton\
hydride resonance during the chemical
reaction~\cite{berner_sambadena_2019,pravdivtsev_transfer_2015,goldman_design_2006,rodin_constant-adiabaticity_2021,vaneeckhaute_long-term_nodate,kiryutin_parahydrogen_2017,barskiy_sabre_2019,knecht_quantitative_2019,knecht_indirect_2019,reineri_hydrogenative-phip_2021,Markelov2021Sep},
and applying a hard $\pi/2$ purge pulse to deplete the $\ket{T_0}$ state prior to
the polarisation transfer
step~\cite{kiryutin_parahydrogen_2017,knecht_efficient_2019,barskiy_sabre_2019,theis_light-sabre_2014,reineri_hydrogenative-phip_2021,natterer_investigating_1998,Markelov2021Sep}.
These two methods are illustrated in Fig.~\ref{fig:rxnscheme}d. 

As we show in the following, the study of ST mixing is 
greatly facilitated by microfluidic PHIP, since instabilities associated
with bubbling experiments are avoided. Additionally,
since hydrogenative PHIP relies on irreversible chemical reactions, the chemical kinetics
influence the observed spectra, and the sample under study would need to be
replaced upon the reaction reaching completion. This is a particular issue
if the samples are scarce or expensive due to isotopic enrichment.
Finally, since hyperpolarised nuclei are in a non-equilibrium state,
the NMR signals relax on a timescale of seconds to tens of seconds, unique to
each molecular species and nuclear spin site, which can convolute the observed
results. This is especially problematic if the signals relax quickly compared
to the time it takes for a shaken tube to be placed in the NMR magnet, or for
bubbles to settle in solution.

\section{Materials and Methods}

All experiments were performed in a 11.7~T magnet using a Bruker AVANCE III
spectrometer system.  The NMR experiments were performed with a custom-built
probe delivering \proton\ RF pulses of 125~kHz amplitude~\cite{sharma_modular_2019}. \proton\ spectra were collected with a 16~ppm
spectral width and 8~k point density.

\textit{Para}-enriched hydrogen gas (gas purity 99.995\%) was continuously
produced by a Bruker parahydrogen generator BPHG90, with a specified
parahydrogen content of 89\%. 

\cbstart
All chemical compounds were purchased from Sigma Aldrich (United Kingdom) and were used as received. All NMR experiments were performed using a precursor solution of 100~mM \ADCA, 6~mM \Cat\ catalyst and 200~mM sodium sulfite dissolved in
D$_2$O at 50\degC. It's been reported that sodium sulfite improves the selectivity of the trans hydrogenation reaction, although the mechanism of its action is not yet known~\cite{Wienands2021Jun,ripka_hyperpolarized_2018}.
\cbend
\subsection{Microfluidic device}

\begin{figure*}[h!]
    \centering
    \includegraphics[width=\linewidth]{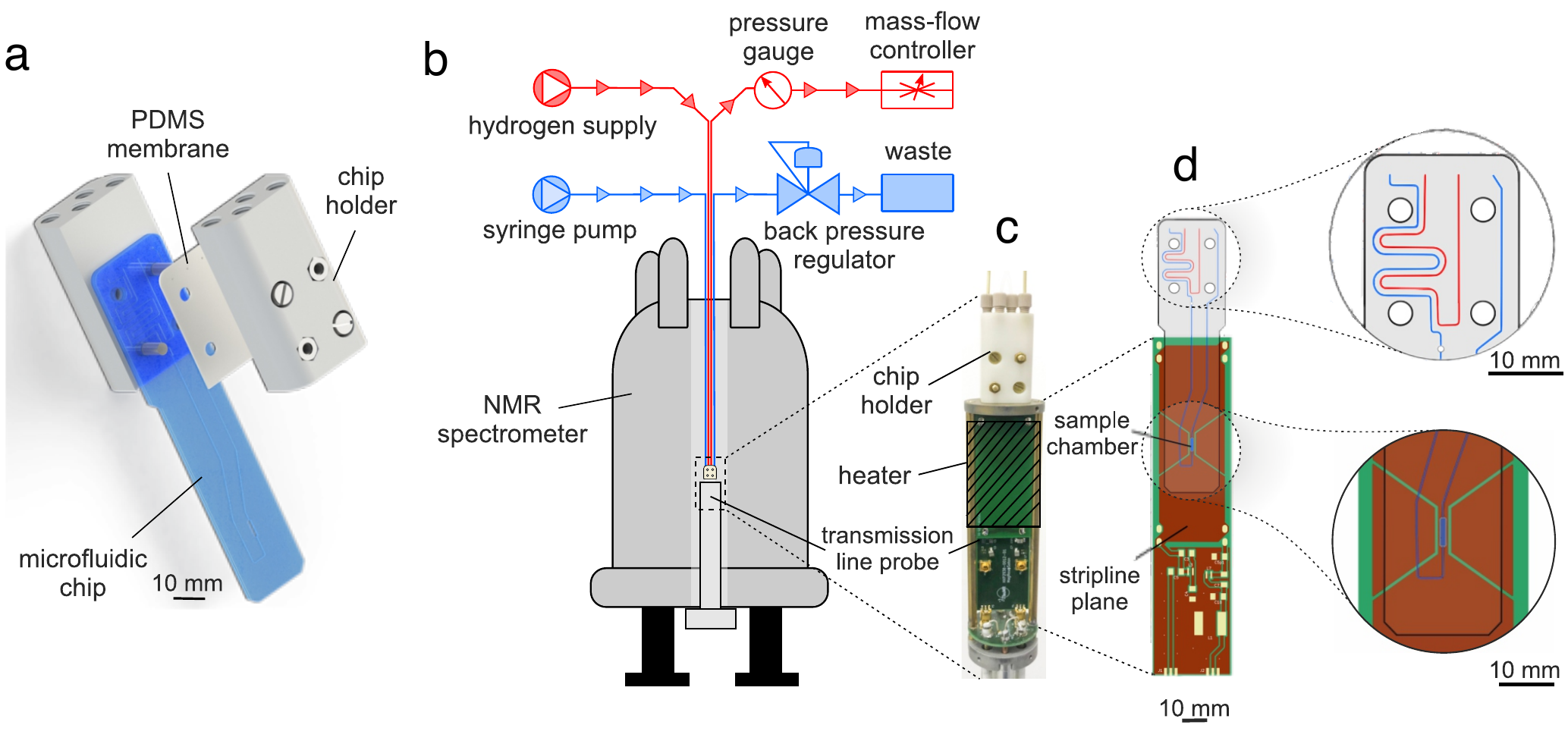}
	\caption{ a) Microfluidic chip assembly. b) A schematic diagram of the
	experimental setup. c) Transmission line probe with the heater indicated as
	the shaded area. d) Drawing of the microfluidic device aligned with the
	stripline plane of the detector. The key areas of the drawing are enlarged.}
    \label{fig:setup}
\end{figure*}

The microfluidic device was made from three layers of polycarbonate (PC) (Self Adhesive Supplies, United Kingdom) with 0.25, 0.5, and
0.25~mm thickness for the top, middle, and bottom layers, respectively. The layers
were cut from PC sheets by a LS3040 $\mathrm{CO_2}$ laser cutter (HPC Laser Ltd, United Kingdom), and
were thermally bonded together as described elsewhere \cite{Yilmaz2016May}. 
A semi-permeable
polydimethylsiloxane (PDMS) membrane of 1~mm thickness (Shielding Solutions, United Kingdom) was placed over the top half of the chip
to seal the channels, but allow hydrogen diffusion from the gas channel to
the liquid channel. The chip and membrane were held together by 3D printed
holders (ProtoLabs, United Kingdom) that attach threaded connectors for 1/16`` capillaries (Cole-Parmer, United Kingdom) to the four access
points on the chip for gas and liquid inlets and outlets. 

In the magnet, the device was placed in a home-built transmission line probe as
shown in \cbstart Fig.~\ref{fig:setup}c. \cbend A heater was clamped outside of the stripline
planes to heat the sample chamber in the microfluidic chip to 50\degC. This is indicated by the shaded area in \cbstart Fig.~\ref{fig:setup}c. \cbend The heated area did not
include the 3D printed holders so that the solution in contact with the
hydrogen gas was kept at lower temperature in order to  maximise the solubility of the
hydrogen gas. The reaction products were detected in a 2.5~$\mathrm{\mu L}$
sample chamber.  The chamber of the chip was aligned with the constrictions of
the stripline detector as shown in Fig.~\ref{fig:setup}d
~\cite{sharma_modular_2019}.  

\subsection{Experimental Procedure}
All experiments were performed inside the high-field NMR spectrometer as shown
in Fig.~\ref{fig:setup}b. Experiments were conducted at 50\degC\ (at the sample chamber only) with the
supply of hydrogen gas set to a pressure of 5~bar and flow rate of
10~$\mathrm{mL~min^{-1}}$, stabilised by a mass flow controller (Cole-Parmer, United Kingdom) connected at
the end of the line. The flow of the precursor solution was controlled with a
syringe pump (Cole-Parmer, United Kingdom) located outside the spectrometer. The target flow rate was set to
10~$\mathrm{\mu L~min^{-1}}$.
Under these operating conditions, the NMR signal reached a steady-state after 10~minutes.

Proton singlet order in [1-\carbon]fumarate was converted into observable
magnetisation using the singlet-to-magnetisation (S2M) \cbstart pulse sequence shown in
Fig.~\ref{fig:rxnscheme}e. \cbend Maximum efficiency was achieved using the following
parameters: $\tau=15.6$~ms, $n_2=14$, $n_1=7$. The repetition delay was set to
60~s.

CW-S2M experiments were performed by applying continuous wave irradiation for 20~s at 0.5 and 2~kHz, while changing the resonance offset from 20 to $-20$~ppm. $\theta$-S2M experiments were performed by applying a hard pulse of varying flip angle prior the S2M pulse sequence. This was achieved by varying the pulse duration from 0 to 8~$\mu s$ in steps of 0.22~$\mu s$.  

\cbstart 
Reference spectrum was obtained using hydrogen in thermal equilibrium. The \proton \, spectrum was obtained by applying a $\frac{\pi}{2}$ pulse and averaging over 400 scans with a recycle delay of 20~s. 
\cbend

\section{Results and Discussion}
\begin{figure}[h!]
    \centering
    \includegraphics[width=\linewidth]{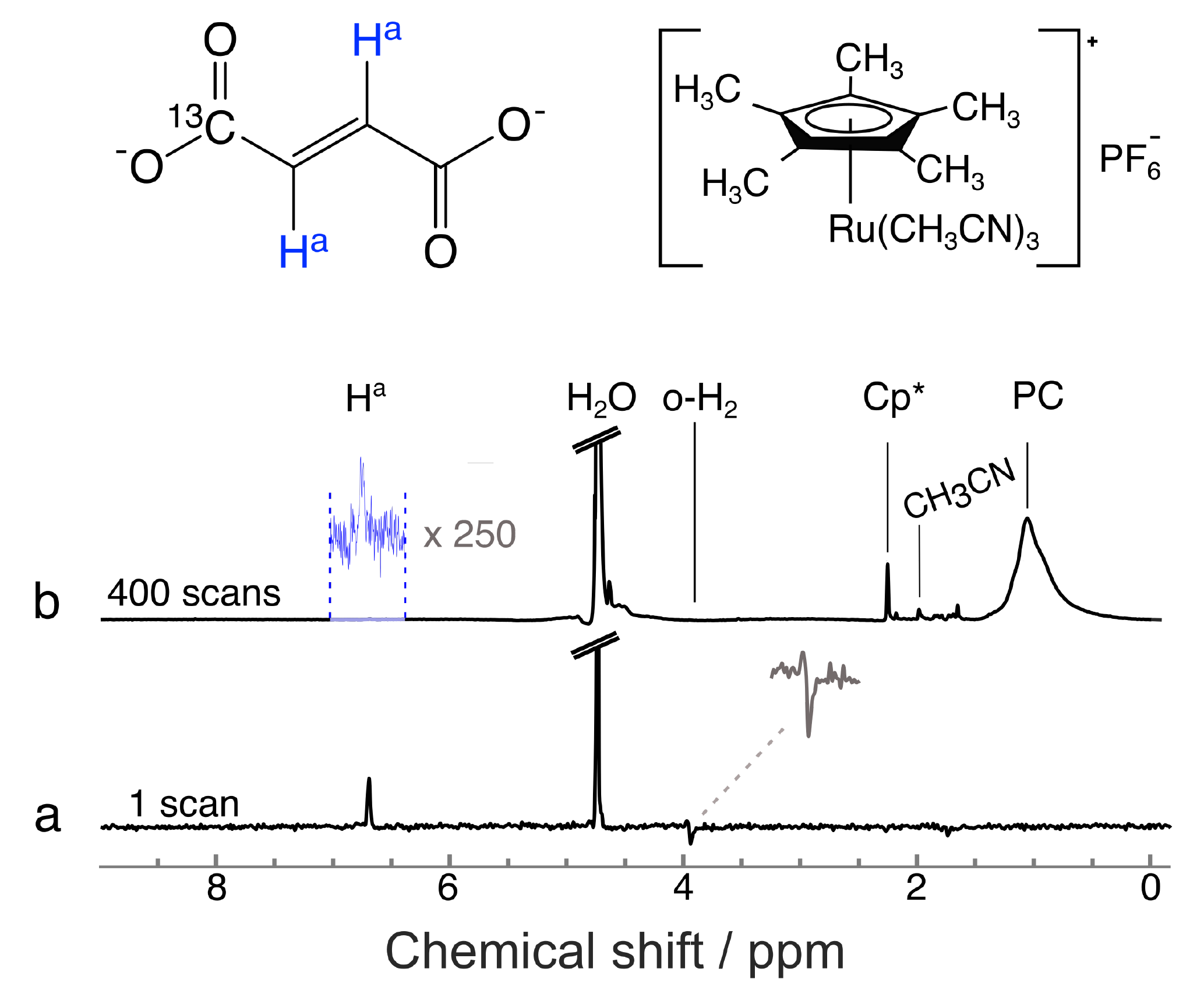}
	\caption{Steady state $\mathrm{^1}$H NMR spectra of [1-\carbon]fumarate sample flowing at 10~$\mathrm{\mu L \,
	min^{-1}}$ in a microfluidic device. a) Spectrum
	collected with the S2M pulse sequence with 89\% parahydrogen. The trace displays a hyperpolarised
	[1-\carbon]fumarate peak at 6.6~ppm. The presence of exchanging hydrogen
	species is indicated at 4~ppm ($\mathrm{o-H_2}$).  b) Reference
    spectrum resulting from a $\pi/2$ pulse with hydrogen in thermal equilibrium. $\mathrm{Cp^\ast}$: catalyst methyl protons,
    PC: background signal from the polycarbonate chip material.}
    \label{fig:enhancement}
\end{figure}

Fig.~\ref{fig:enhancement}a depicts a single-scan proton NMR spectrum obtained
after application of the S2M pulse sequence in a steady-state flow experiment with
89\% para-enriched $\mathrm{{H_2}}$.
This can be compared to the 400-scan reference spectrum obtained
after application of a $\pi/2$ pulse using hydrogen in thermal equilibrium  (i.e., not
\emph{para}-enriched) in Fig.~\ref{fig:enhancement}b.

The spectra contain a
peak at 6.6~ppm that corresponds to the fumarate protons $\mathrm{H^a}$.
From the ratio of the signal intensity in the reference and hyperpolarised spectra, the $\mathrm{^1H}$ polarisation was estimated. Accounting for the difference in the number of scans, the signal enhancement was calculated as $190\pm 10$. At the field of 11.7~T and temperature of 50\degC\, this
corresponds to $0.7\pm 0.1$\% $\mathrm{^1H}$ polarisation. At 10 $\mathrm{\mu L \, min^{-1}}$ flow rate, the concentration of fumarate was 1.2 $\pm$ 0.5 mM , which corresponds to  1.2 $\pm$ 0.5 \% yield. This was calculated by comparing the intensity of the $\mathrm{Cp^*}$ peak in the reference spectrum to the intensity of the fumarate peak and accounting for the difference in the number of protons. 

\cbstart
\cbend

\begin{figure}[ht!]
    \centering
    \includegraphics[width=0.8\linewidth]{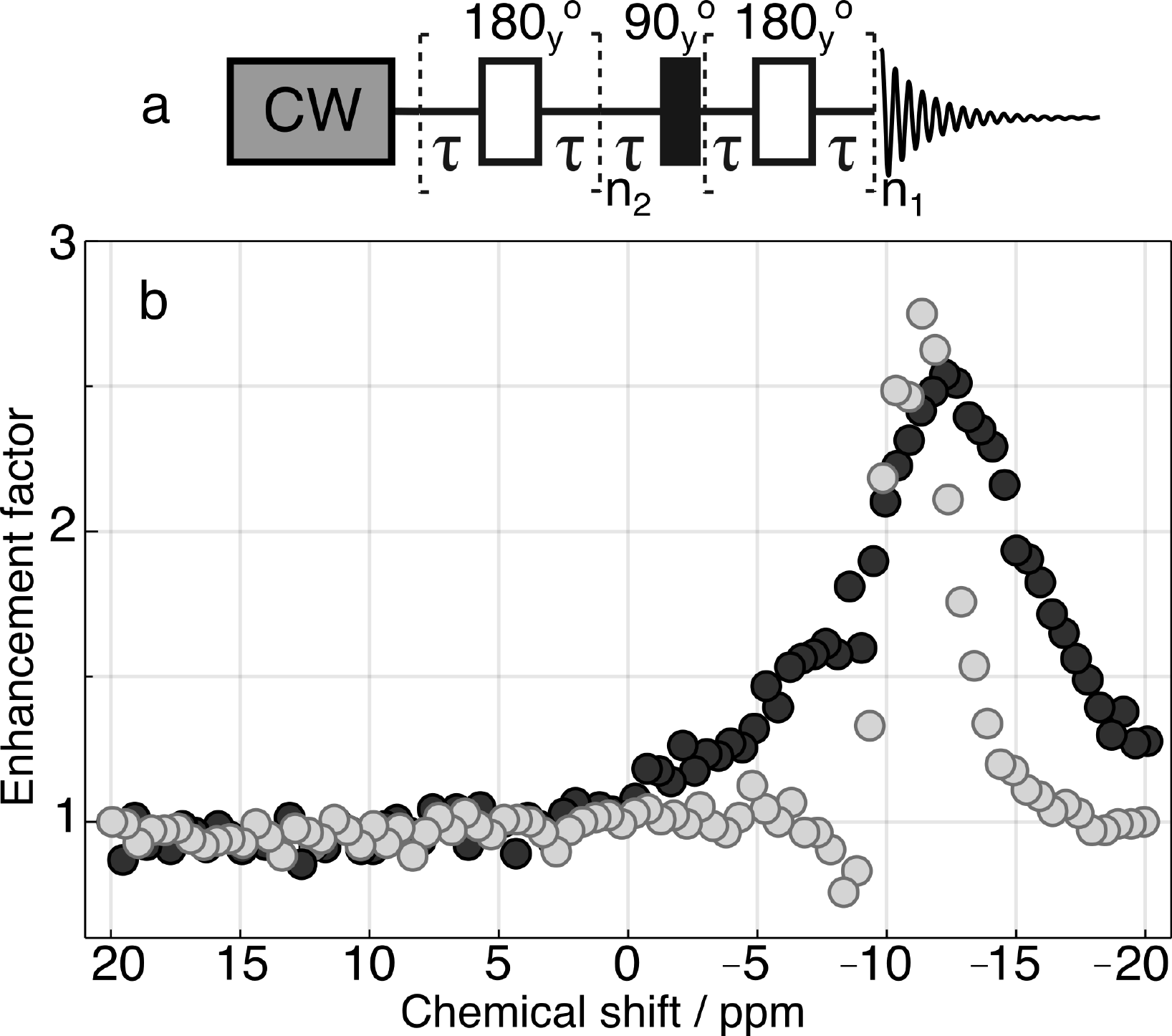}
    \cbstart
	\caption{a) Singlet-to-magnetisation pulse sequence with spin-locking field
	applied during recycle delay.  b) Integral of signal intensity of the hyperpolarised
	proton of [1-\carbon]disodium fumarate as a function of the resonance offset of
	the spin-locking field. Experiments were preformed with two CW amplitudes, corresponding to 2~kHz and 0.5~kHz nutation frequency shown as black and grey data points, respectively.
	Signal amplitude was normalised to the signal acquired with CW frequency set to 20 ppm.
	}
	\cbend
    \label{fig:CWS2Mresults}
\end{figure}
\cbstart The hyperpolarised spectrum features the aforementioned partially negative line
at 4~ppm labelled 
$\mathrm{o-H_2}$. \cbend The heavy metal catalyst and dissolved molecular hydrogen form intermediate complexes where the two hydrogen nuclei occupy chemically inequivalent positions. At high magnetic field this introduces a chemical shift difference between the two protons, which causes singlet state population to leak into the population of the central triplet state.
In addition, the chemical shift difference lifts the degeneracy of the two triplet state transitions. In rapid exchange, this leads to a small partially negative line in the dissolved $\mathrm{H_2}$ signal~\cite{kiryutin_parahydrogen_2017,knecht_mechanism_2018,knecht_indirect_2019,knecht_efficient_2019}, as displayed in the spectrum in Fig.~\ref{fig:enhancement}a.

To suppress the effects of ST mixing we performed experiments in which
we applied continuous-wave (CW) irradiation to the sample for 20~s prior to
the application of S2M and signal acquisition. The pulse sequence is shown in \cbstart Fig.~\ref{fig:CWS2Mresults}a. \cbend
The resulting integral of fumarate signal intensity at 6.6~ppm is plotted as a function of
CW offset frequency in Fig.~\ref{fig:CWS2Mresults}b. Experiments were performed with two different CW 
amplitudes, corresponding to
0.5 kHz and 2 kHz nutation frequency on protons shown as grey and black circles, respectively. 

The profiles of signal intensity against the CW irradiation frequency display a
peak at around $-11$~ppm. \cbstart This is a typical chemical shift of hydride
species for ruthenium complexes~\cite{rosal_dft_2008}, indicating that ST
mixing does indeed occur for the hydride species, and is suppressed by CW
irradiation. The \proton\, spectra can be used to observe ST mixing and this has been shown in case of SABRE by either applying a single hard pulse after CW irradiation or a pulse sequence designed to probe higher spin-order if hydride species undergo very fast chemical exchange \cite{knecht_mechanism_2018,kiryutin_parahydrogen_2017}. In the present case, hydride species are not directly observable due to fast exchange and low sensitivity. \cbend The signal is enhanced by a factor of $\sim$3 when the
spin-locking amplitude is set to either to 0.5~kHz or 2~kHz applied at $-11$~ppm. The peak width in each case corresponds
roughly to the excitation bandwidth, resulting in a narrower peak at the lower CW amplitude.

\cbstart
\begin{figure}[h]
    \centering
    \includegraphics[width=\linewidth]{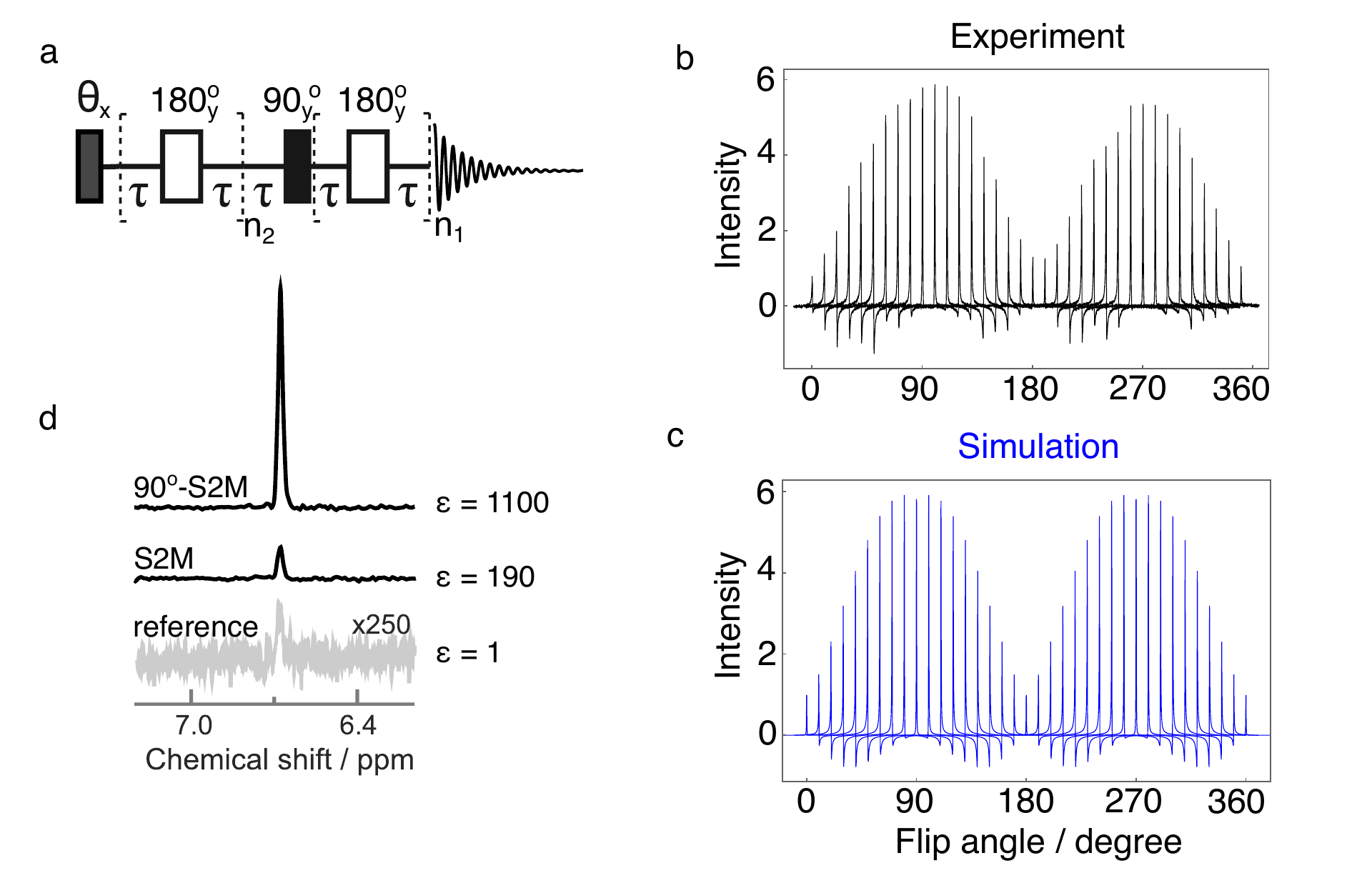}
    \caption{a) $\theta$ - S2M pulse sequence. The $\theta$ angle was arrayed from $0^\circ$ to $360^\circ$ in steps of $10^\circ$. $\tau = 15.6$~ms, $n_1$ = 7, $n_2$ = 14. b) Experimentally obtained $\mathrm{H^a}$ signals of [1-\carbon]fumarate as a function of the purge pulse angle. The y-axis shows the improvement of the enhancement factor compared to S2M without the purge pulse. c) Computational simulation of the spin system using SpinDynamica software~\cite{Bengs2018Jun}. d) Comparison of the signal intensity of fumarate protons ($\mathrm{H^{a}}$) between the reference spectrum, pure S2M and $90^\circ - \mathrm{S2M}$.}
    \label{fig:90S2Mresults}
\end{figure}
\cbend

We contrast this to another method, which has been used to address ST mixing
effects: applying a hard pulse (which we will refer to as the purge pulse) to the protons prior to polarisation transfer
and signal
acquisition. Application of a $\pi /2$ purge pulse on the proton channel depletes the $\ket{T_0}$ state, which partially reconstitutes the population difference between the $\ket{S_0}$ and $\ket{T_0}$ states
~\cite{kiryutin_parahydrogen_2017,knecht_efficient_2019,barskiy_sabre_2019,theis_light-sabre_2014,reineri_hydrogenative-phip_2021,natterer_investigating_1998}.

Fig.~\ref{fig:90S2Mresults}a shows the pulse sequence used to investigate the
phenomenon, and Fig.~\ref{fig:90S2Mresults}b shows the hyperpolarised $\mathrm{H^a}$ proton signals obtained experimentally by varying the flip angle $\theta$. 
The flip angle
was varied from $0^\circ$ to $360^\circ$ in steps of $10^\circ$. The signal shows an oscillatory dependence 
on the flip angle of the purge pulse, with maxima occurring at
$90^\circ$ and $270^\circ$, and no improvement seen near $180^\circ$. The signal at $270^\circ$ is about 15\% less than at $90^\circ$. \cbstart While this is likely due to $B_1$ inhomogeneities, other factors, for example chemical kinetics on the timescale of the pulse length, may also contribute. \cbend

The spectral peaks in \cbstart Fig.~\ref{fig:90S2Mresults} \cbend also display phase distortions depending on the flip angle of the purge pulse. The origin of the effect was confirmed by numerical simulations using software package SpinDynamica  \cite{Bengs2018Jun} and the result is shown in \cbstart Fig.~\ref{fig:90S2Mresults}c. \cbend
The simulation assumes that before the application of the sequence depicted in \cbstart Fig.~\ref{fig:90S2Mresults}a, \cbend the $\ket{\mathrm{S_0}}$ and  $\ket{\mathrm{T_0}}$ states are 55\%  and 45\% populated, respectively.
The populations of the other triplet states are neglected.
The agreement between experimental data and numerical simulation is striking.
\cbstart Both experiments and simulations show phase distortions when the flip angle is not an integer multiple of 90 degrees. These phase distortions arise as follows: When the first pulse has a flip angle of 90 degrees, the pulse transfers the population of the central triplet state $\ket{T_0}$ to the outer triplet states $\ket{T_{\pm 1}}$, increasing the population difference between the singlet state $\ket{S_0}$ and the central triplet state $\ket{T_0}$, and hence enhancing the hyperpolarised NMR signal at the end of the pulse sequence. However, the flip angle of the first pulse is not a multiple of 90 degrees, the transport of populations between the triplet state is accompanied by the excitation of single-quantum triplet-triplet coherences, of the form $\ket{T_{\pm 1}} \bra{T_0}$ and $\ket{T_0} \bra{T_{\pm 1}}$. These coherences persist throughout the pulse sequence, and appear as out-of-phase signal components in the observed spectrum, which have the effect of an undesirable phase shift of the observed peak. \cbend

In Fig.~\ref{fig:90S2Mresults}d a comparison is shown between the
reference spectrum obtained with 400 scans and single scan NMR spectra of
protons $\mathrm{H^a}$ after applying the S2M sequence with a purge pulse of
0$^\circ$ and $90^\circ$. The enhancement in the latter case was
calculated to be $1100\pm10$ in contrast to $190\pm10$ without applying the purge pulse. This corresponds to $4.0\pm 0.1\%$ \proton\ polarisation, and hence a nearly
6-fold improvement in achievable fumarate signal. The enhancement factor was calculated by comparing the integral of the $\mathrm{H^a}$ peak in the hyperpolarised and reference spectrum, accounting for the difference in the number of scans between the two spectra.

The hard-pulse method yielded a 6-fold improvement in the achievable fumarate signal, compared to 3 for the spin-locking method. This was unexpected since the
spin-locking method can in principle lead to higher signal enhancements as it should mitigate the effect of ST mixing entirely. We
believe the lower efficiency provided by spin locking is due to the micro-NMR
probe design where the RF field is concentrated exclusively onto the sample chamber as shown
in Fig.~\ref{fig:setup}d. Therefore the solution outside the sample chamber is
not affected by the RF irradiation, and thus the ST mixing cannot be suppressed
for molecules of fumarate that formed in the channels before
reaching the sample chamber. This is not a problem for the hard-pulse method
since the pulse is applied after the chemical reaction.

The results obtained show the remarkable reproducibility and stability of the
chemical reactions performed in the microfluidic device over the course of
hours.  A steady-state between the rate of chemical reaction to form the
hyperpolarised product and the rate of relaxation was established, and without
the confounding influence of these external factors it is possible to study and
optimise pulse sequences in hyperpolarised NMR experiments.  An additional
benefit of working on a microfluidic scale is the small sample volumes
required, meaning expensive or scarce samples can be more readily used. For
example, the data in Fig.~\ref{fig:90S2Mresults}b required 40~minutes of
experimental time, consuming 400 $\mathrm{\mu L}$ of solution, which is the
approximate volume required for a single PHIP experiment in a conventional 5~mm
NMR tube.

The yield of fumarate in the chip was 1.2 $\pm$ 0.5 \%. The low yield of the reaction is
most likely due to the limited uptake of hydrogen into the flowing solution. Finite
element simulations of the chip have shown that when methanol is flowed through
the chip at $\mathrm{10 \, \mu L \, min^{-1}}$ at pressure of 5 bar, only 10 mM
of hydrogen dissolves in the fluid \cite{ostrowskaSpatiallyResolvedKinetic2021}. Since in this work water was used as the solvent, the concentration of hydrogen dissolved is expected to be lower
due to poorer solubility of hydrogen in water. \cbstart Modifications to the apparatus to improve the $\mathrm{H_2}$ uptake and yield of the reaction are currently underway. \cbend

\section{Conclusion} 
\cbstart 

In this work we employed a microfluidic chip to run PHIP reactions,
incorporating the hydrogenation, sample transport, RF excitation and signal
detection steps onto a single device. In the reaction we hyperpolarised
[1-\carbon]fumarate, and used the S2M pulse sequence to generate in-phase
proton magnetisation for observation in the 2.5~$\mu$L sample chamber, achieving 4\% proton polarisation. We used
this system to investigate pulsed NMR methods that reduce the detrimental
effects of singlet-triplet mixing in this PHIP reaction. We showed that
application of continuous wave irradiation prior to applying the S2M pulse
sequence leads to a 3-fold improvement to the fumarate proton polarisation, and
also allowed us to locate the chemical shift of the catalyst complex on which
singlet-triplet mixing occurs. We contrasted that with application of a $\pi$/2
pulse prior to applying the S2M sequence, which led to a 6-fold improvement to
the proton polarisation.

This continuous-flow PHIP approach allows one to establish a constant stream of
a hyperpolarised product, providing stable and reproducible conditions for the
study of complex chemical and spin-dynamical phenomena in a well-controlled
environment. This
is an important step towards
observation of metabolism in biological systems by hyperpolarised NMR 
on a single microfluidic device. By bringing hydrogen gas into solution through a membrane as
opposed to bubbling or shaking, the chemical reaction is more stable and
reaches a steady-state with a variation in the concentration of reaction
product of 1\%. By operating at a small volume-scale (microliters), the
consumption of expensive materials is significantly reduced as compared to
performing reactions in NMR tubes. 

Not only does microfluidic implementation aid in the development of
hyperpolarised NMR methods, but incorporating hyperpolarisation to enhance NMR
signals opens the door to the use of NMR as a detection method to study
biological systems in microfluidic devices. Methods such as fluorescence
spectroscopy require using specific fluorescent tags to track molecules, and
UV-visible spectroscopy offers limited ability to identify molecules. The
molecular specificity and non-destructive nature of NMR spectroscopy makes it
an ideal technique to track metabolic reactions, and direct production of
hyperpolarised fumarate in a microfluidic chip is an important step towards
this goal. However, further developments are required to make this dream a
reality, such as the removal of toxic chemicals after the hyperpolarisation
process, and the incorporation of \carbon\ NMR for background-free detection
with high chemical specificity and resolution. 

Much work with hyperpolarised biomolecules relies on \carbon\, hyperpolarisation and
detection, since this is preferable for \emph{in vivo} imaging as the large
background signals from water molecules are not present. The probe used for
this work is doubly tuned for \proton\, and \carbon\, excitation and
detection but in order to perform such experiments several issues need to be addressed. A
prerequisite of using PHIP-polarised metabolites for biological studies is the
ability to remove the catalyst and reaction side-products from the solution.
This has been shown to be possible for [1-\carbon]fumarate via a precipitation
procedure~\cite{Knecht2021Mar}, and for a
variety of other PHIP-polarised metabolites via the side-arm hydrogenation
procedure~\cite{reineri_hydrogenative-phip_2021}. Precipitation procedures are not feasible in microfluidic devices as the solid would block the fluidic channels. However, scavenger compounds that bind the catalyst could potentially be used for this purpose \cite{Barskiy2018Jun,Kidd2018Jul}.

\cbend
\begin{acknowledgement}
The authors are indebted to Dr.~Christian Bengs and Dr.~Giuseppe Pileio for help with the SpinDynamica simulations,
as well as for insightful discussions about the results and theory.
This work has been supported by an EPSRC iCASE studentship EP/R513325/1 to SJB, co-funded by Bruker UK Ltd.,
as well as by the EU H2020 FETOpen Project "TISuMR" (Grant number 737043). This project has also received funding from the European Union's Horizon 2020 research and innovation programme under the Marie Skłodowska-Curie Grant Agreement No. 766402, as well as  ERC project 786707-FunMagResBeacons, and EPSRC grants EP/P009980/1 and EP/V055593/1.
\end{acknowledgement}

\bibliography{fumarate-library}

\end{document}